\documentclass[12pt]{ronbun}

\usepackage{amsmath,amssymb}
\usepackage{graphicx,color}
\usepackage{cite}
\usepackage{bm}
\usepackage{dcolumn}

\newcommand{\nn}{\nonumber}

\newcommand{\del}{\delta}

\newcommand{\al}{\alpha}

\renewcommand{\th}{\theta}

\numberwithin{equation}{section}

\begin{document}

\begin{flushright}
\parbox{4.2cm}
{OCU-PHYS-208 \hfill \\
KEK-TH-948 \hfill
{\tt hep-th/0403243}
 }
\end{flushright}

\vspace*{1.1cm}

\begin{center}
 \Large\bf Notes on D-branes of Type IIB String on AdS$_5$ $\times$ S$^5$
\end{center}
\vspace*{1.5cm}
\centerline{\large Makoto Sakaguchi$^{\dagger a}$ and Kentaroh
Yoshida$^{\ast b}$}
\begin{center}
$^{\dagger}$\emph{
Osaka City University Advanced Mathematical Institute (OCAMI)\\
Sumiyoshi, Osaka 558-8585, Japan.
}\vspace*{0.5cm}
\\
$^{\ast}$\emph{Theory Division, High Energy Accelerator Research 
Organization (KEK),\\
Tsukuba, Ibaraki 305-0801, Japan.} 
\\
\vspace*{1cm}
$^{a}$msakaguc@sci.osaka-cu.ac.jp
~~~~
$^{b}$kyoshida@post.kek.jp
\end{center}

\vspace*{1cm}

\centerline{\bf Abstract}

\vspace*{0.5cm}

We promote a study of D-branes of type IIB string on the AdS$_5 \times
S^5$ background. The possible D-branes preserving half of
supersymmetries were classified up to
and including the fourth order of fermionic variable $\th$ in our
previous work [hep-th/0310228]. In this paper we show that our
classification is still valid even at the full order of $\th$\,. This
proof supplements our previous results and completes the classification
of D-branes in the type IIB string theory on the AdS$_5\times S^5$\,.
  
\vspace*{0.5cm}

\vfill

\thispagestyle{empty}
\setcounter{page}{0}

\newpage 

\section{Introduction}

D-brane is an important key ingredient in studies of non-perturbative
aspects of superstring theories \cite{Pol}.  A recent interest is to
investigate D-branes on general curved backgrounds, motivated by recent
developments in studies of pp-wave backgrounds.  To begin with, the
maximally supersymmetric type IIB pp-wave background was found
\cite{BFHP1}. Then the Green-Schwarz type IIB string theory on this
pp-wave was shown to be exactly solvable in the light-cone
gauge\cite{M,MT}.  After that D-branes on the pp-wave were intensively
investigated \cite{SMT,BP,DP,Bain,BGG,SMT2} since one can study directly
them by solving classical equations of motion and quantizing the theory.

The covariant studies of D-branes in type IIB and IIA strings on
pp-waves were discussed in \cite{BPZ} and \cite{HPS}, respectively, by
using the method of Lambert and West \cite{LW}.  Motivated by these
developments, we have carried out the covariant analysis for D-branes of
type IIB string on the AdS$_5\times S^5$ background \cite{SaYo4}. The
allowed 1/2 supersymmetric (SUSY) D-brane configurations have been
classified.  Our result is also consistent to that of brane probe
analysis done in \cite{SMT}.  In addition, Penrose limit
\cite{P,BFHP2}\footnote{Penrose limit of superalgebra in the
AdS$_5\times S^5$ background was discussed in \cite{HKS:10}.} of D-branes
in the AdS$_5\times S^5$ has been discussed and we have seen that the
result after this limit agrees with the possible D-brane configurations
in the type IIB pp-wave background.  On the other hand, by combining the
methods proposed in \cite{EMM,dWPP}, the covariant method is also
applicable to open supermembrane theory on the pp-wave
\cite{SY1,SaYo:pp} and AdS$_{4/7}\times S^{7/4}$ \cite{SaYo:ads}
backgrounds. These results are related via Penrose limit and are also
consistent to the brane probe analysis in eleven dimensions
\cite{Kim-Yee}.

In this paper, we continue to study D-branes of type IIB string on the
AdS$_5\times S^5$ background \cite{SaYo4}. The previous analysis at the
fourth order of $\th$ is extended to the full order. We show that the
higher order terms with respect to $\th$ do not affect the
classification of 1/2 SUSY D-branes at the fourth order under the
conditions obtained in the fourth order analysis. This proof completes
our classification of D-branes.

The organization of the present paper is as follows: In section 2, we
introduce our previous result on the classification of D-branes in type
IIB string theory on the AdS$_5\times S^5$ background, based on the
analysis up to and including the fourth order of $\th$\,. In section 3,
we show that the classification is still valid even at the full order of
$\th$\,.  Section 4 is devoted to a conclusion and discussions.

\section{Classification of 1/2 SUSY D-branes}

Here we will briefly review our classification result of 1/2 SUSY
D-branes of type IIB string theory on the AdS$_5\times S^5$ background
\cite{SaYo4}. We work in the notation and convention used in
\cite{SaYo4}.

The open-string world-sheet $\Sigma$ has the one-dimensional boundary
$\partial\Sigma$\,, and we can impose the Neumann and Dirichlet boundary
conditions on $\partial\Sigma$\,.  These conditions are represented by
\begin{eqnarray}
&& \partial_{\sigma}X^{\overline{A}} \equiv \partial_{\sigma}X^M
e_M^{\overline{A}} = 0 \qquad (\mbox{Neumann condition})\,, \\ &&
\partial_{\tau}X^{\underline{A}} \equiv \partial_{\tau}X^M
e_M^{\underline{A}} = 0 \qquad (\mbox{Dirichlet condition})\,,
\end{eqnarray}
where we have used the overline as $\overline{A}_i~(i=0,\dots,p)$ for
the indices of Neumann coordinates and the underline as
$\underline{A}_j~(j=p+1,\dots,9)$ for the indices of Dirichlet
coordinates. 

By using the projection operator $P^\pm$, a boundary condition is
imposed on the fermionic variable $\th$ as
\begin{eqnarray}
P^{\pm}\th = \th\,, \qquad  P^{\pm} = \frac{1}{2}(1\pm M)\,. 
\end{eqnarray}
The gluing matrix $M$ is described as follows: 
\begin{eqnarray}
\label{cond1}
&& M = \left\{
\begin{array}{l}
m\otimes i\sigma_2\,, \quad d=2~(\mbox{mod}~4)\quad p=-1,3,7 
\\
m\otimes \rho\,, \quad d=4~(\mbox{mod}~4)\quad p=1,5,9
\end{array}
\right.\,, \\
&& m = s\Gamma^{\underline{A}_1}\cdots\Gamma^{\underline{A}_d}\,, \quad 
 s = \left\{
\begin{array}{l}
1 \quad \mbox{for $X^0$: Neumann} \\ 
i \quad \mbox{\,for $X^0$: Dirichlet}
\end{array}
\right.\,,~~
\rho=\left\{
  \begin{array}{ll}
   \sigma_1    & \mbox{when}~~ \sigma=\sigma_3    \\
   \sigma_3    & \mbox{when}~~ \sigma=\sigma_1   \\
  \end{array}
\right.\,.
\nonumber
\end{eqnarray}
In \cite{SaYo4}, a classification of 1/2 SUSY D-branes in the
AdS$_5\times S^5$ was given by considering the vanishing conditions of
the $\kappa$-variation surface terms up to and including the fourth
order in $\theta$.  

For the $d=2$ (mod 4) case, the possible
configurations of D-branes need to satisfy the following condition:
\begin{itemize}
 \item The number of Dirichlet directions in the AdS$_5$ coordinates
       $(X^0,\cdots,X^4)$ is even, and the same condition is also satisfied for
       the $S^5$ coordinates $(X^5,\cdots,X^9)$.
\end{itemize}
For the $d=4~({\rm mod}~4)$ case, 
D-branes satisfying the following condition are allowed:  
\begin{itemize}
 \item The number of Dirichlet directions in the AdS$_5$ coordinates
       $(X^0,\cdots,X^4)$ is odd, and the same condition is also satisfied for
       the $S^5$ coordinates $(X^5,\cdots,X^9)$.
\end{itemize}
The D-branes in the AdS$_5\times S^5$ are restricted with respect to the
directions to which a brane world-volume can extend. All of possible
D-brane configurations at the origin are summarized in Tab.\,\ref{tab1}.
\begin{table}[htbp]
 \begin{center}
  \begin{tabular}{|c|c|c|c|c|c|}
\hline
D-instanton & D-string  & D3-brane  & D5-brane & D7-brane & D9-brane \\
\hline\hline
(0,0) & (0,2),~(2,0) & (1,3),~(3,1) & (2,4),~(4,2)
& (3,5),~(5,3) & absent \\
\hline
  \end{tabular}
 \end{center}
\caption{The possible 1/2 supersymmetric D-branes in AdS$_5\times S^5$
sitting at the origin.}
\label{tab1}
\end{table}
When we consider the D-branes sitting outside the origin, 
only a D-instanton is allowed. 

In the next section we will discuss that the higher order terms with
respect to $\theta$ do not modify the above classification of D-branes
sitting at and outside the origin.

\section{Validity of the Classification at Full Order of $\th$}

Now let us show that our classification at the fourth order
of $\th$ is still valid at the full order of $\th$\,. 
We shall start from the covariant Wess-Zumino term \cite{MT:action}:
\begin{eqnarray}
{\cal L}_{\rm WZ} = -2i\int^1_0\!\!dt\,\widehat{E}^A\bar{\th}\Gamma_A\sigma
\widehat{E}\,, 
\end{eqnarray}
where $\widehat{E}^A(X,\th) \equiv E^A(X,t\th)$ and
$\widehat{E}^{\al}(X,\th) \equiv E^{\al}(X,t\th)$\,. 

We notice that the surface term coming from the $\kappa$-variation 
is represented by\footnote{We will omit the symbol ``hat'' of 
$E^A$ and $E^{\al}$ because 
the shift $\th \rightarrow t\th$ 
does not affect our discussion. } 
\begin{eqnarray}
E_\tau^A(\bar{\th}\Gamma_A)_{\al}\sigma \del_{\kappa}Z^{\widehat{M}}
E_{\widehat{M}}^{\al}\,,
\label{surface}
\end{eqnarray}
where $E_{\tau}^A$ denotes the $\tau$-component of $E_i^A$\,. Here we
should remark that the Nambu-Goto or Dirac-Born-Infeld part of the
action does not produce any surface term under the $\kappa$-variation.
Then, in order to check our classification of D-branes 
at full order of $\th$, it is sufficient to show the key relations:
\begin{eqnarray}
\label{key1}
&& E_{\tau}^{\underline{A}} =
 \partial_{\tau}Z^{\widehat{M}}E^{\underline{A}}_{\widehat{M}} = 0\,,  \\
&& P^{\mp}\del_{\kappa}Z^{\widehat{M}} E_{\widehat{M}}^{\al} = 0 
\qquad \mbox{for}\quad\th = P^{\pm}\th\,
\label{key2}
\end{eqnarray}
under the conditions denoted in Section 2. When the relations
(\ref{key1}) and (\ref{key2}) are proven, we can easily see that the
surface term (\ref{surface}) should vanish as
\begin{eqnarray}
E_{\tau}^{\overline{A}}\bar{\th} P^{\pm}\Gamma_{\overline{A}}\sigma
\del_{\kappa}Z^{\widehat{M}}E_{\widehat{M}} = 
E_{\tau}^{\overline{A}}\bar{\th} \Gamma_{\overline{A}}\sigma
P^{\mp}\del_{\kappa}Z^{\widehat{M}}E_{\widehat{M}} = 0\,.
\end{eqnarray}
Therefore, all we have to do in order to show the validity at full order
is to prove two relations (\ref{key1}) and (\ref{key2})\,.  We will
prove (\ref{key1}) and (\ref{key2}) below by using the 1/2 SUSY
conditions obtained at the fourth order analysis. Before going to the
detail analysis, we should remark about the D-instanton case. 
This case has no Neumann coordinates, and so it is sufficient to see
(\ref{key1}) only in order for the surface term to vanish.

\subsubsection*{Proof of (\ref{key1})} 

Here let us show the relation (\ref{key1}).  For this purpose, we
consider the term $\partial_{\tau} X^M 
E^{\underline{A}}_{M}$ 
and rewrite it as 
\begin{eqnarray}
&& \partial_{\tau}X^M
E^{\underline{A}}_{M} 
= \partial_{\tau}X^M\left(
e_M^{\underline{A}} +
i\bar{\th}\Gamma^{\underline{A}}\left(\frac{\sinh(\mathcal{M}/2)}{\mathcal{M}/2}\right)^2 [D\th]_{M}
\right) \nn \\
 && \qquad \qquad \,\, = \partial_{\tau}X^M\left(
i\bar{\th}\Gamma^{\underline{A}}P^{\mp}
\left(\frac{\sinh(\mathcal{M}/2)}{\mathcal{M}/2}\right)^2 P^{\mp} 
[D\th]_{M}
\right) \nn \\
&& \qquad\qquad \,\, = i\frac{\lambda}{2}\partial_{\tau}X^M e_M^{\overline{B}}
\bar{\th}\Gamma^{\underline{A}}P^{\mp}
\left(\frac{\sinh(\mathcal{M}/2)}{\mathcal{M}/2}\right)^2 P^{\mp}
\widehat{\Gamma}_{\overline{B}}i\sigma_2 \th \nn \\
&& \qquad \qquad\,\, \quad + \frac{i}{4}\partial_{\tau}X^M e_M^{\overline{D}}
\bar{\th}\Gamma^{\underline{A}}P^{\mp}
\left(\frac{\sinh(\mathcal{M}/2)}{\mathcal{M}/2}\right)^2 P^{\mp}
\Gamma_{\overline{B}\underline{C}}\th \omega^{\overline{B}
\underline{C}}_{\overline{D}}\,, 
\label{last}
\end{eqnarray}
where we have introduced the following notation: 
\[
 [D\th]_{M} \equiv \frac{\lambda}{2}e^B_M\widehat{\Gamma}_Bi\sigma_2\th 
+ \frac{1}{4}\omega^{AB}_M\Gamma_{AB}\th\,.  
\]
When we move from the first line to the second line, 
we have used
the
definition of the Dirichlet boundary condition
\begin{equation}
\partial_{\tau}X^M e_M^{\underline{A}} = 0\,,
\end{equation}
and the identities  
\begin{eqnarray}
\label{PMP}
P^{+}\mathcal{M}^{2n}P^{-} =
P^{-}\mathcal{M}^{2n}P^{+} = 0\,, 
\end{eqnarray}
which hold under the conditions obtained in the fourth order analysis.  

The first term in the most right hand side of (\ref{last}) vanishes
under the conditions: the even number of Dirichlet directions are
contained in the case of $d=2$ (mod 4), or the odd number of Dirichlet
ones are in the $d=4$ (mod 4) case. The second term vanishes at the
origin because the spin connection vanishes at the origin.  That is, the
last line of (\ref{last}) vanishes at the origin under the conditions
obtained in the fourth order analysis.  Therefore, we obtain
\begin{eqnarray}
\partial_{\tau}X^M E^{\underline{A}}_{M} = 0\,. 
\end{eqnarray}
In addition, we can easily show the relation
\begin{eqnarray}
\partial_{\tau}\th^{\al} E_{\al}^{\underline{A}} = 0\,,
\end{eqnarray}
under the fourth order conditions. 
Thus, we have shown that the relation (\ref{key1}) should be satisfied 
under the conditions found at the fourth order. 

We should note that (\ref{last}) is trivially zero for the D-instanton
because it contains no Neumann directions. The condition
(\ref{last})$=0$ is satisfied at and outside the origin.  Namely, we
have seen that 1/2 SUSY D-instanton has no modification from
higher order terms of $\th$\,.  And the consideration for 1/2
SUSY D-branes sitting outside the origin has been completed.

Finally, we would like to comment on the physical interpretation of the
condition (\ref{key1})\,.  The condition (\ref{key1}) implies that the
configurations of D-branes are static, because it represents that the
momenta for Dirichlet directions are zero. It would be also helpful to
consider a flat limit ($\lambda\rightarrow 0$) and to see
\begin{eqnarray}
E^{\underline{A}}_{\tau} = \dot{X}^{\underline{A}} -i\bar{\th}
\Gamma^{\underline{A}}\dot{\th} = 0\,. 
\end{eqnarray}
This interpretation should be plausible because D-branes moving in the
target space would be non-supersymmetric rather than 1/2 supersymmetric
ones. It is well known that the $\kappa$-symmetry in open string
theories governs the dynamics of D-branes \cite{CS}.  This fact may be
partially realized in our case as the condition (\ref{key1})\,.

\subsubsection*{Proof of (\ref{key2})}

We shall prove the relation (\ref{key2}) below.  
In order to show it,
we need to prove the relation: 
\begin{eqnarray}
\label{ho}
&& \del_{\kappa}X^M e_M^{\underline{A}} = 0\,. 
\end{eqnarray} 
By definition of the $\kappa$-transformation, we have
\begin{eqnarray}
\del_{\kappa}E^A = \del_{\kappa}X^M E_M^{A} + \del_{\kappa}\th^{\al}E_{\al}^A
=0\,. 
\end{eqnarray}
From this equation, we obtain 
\begin{eqnarray}
\del_{\kappa}X^Me_M^A &=& - i\bar{\th}\Gamma^A
\left(\frac{\sinh(\mathcal{M}/2)}{\mathcal{M}/2}\right)^2 
[D\th]_N e_{B}^{N} e_M^B \del_{\kappa} X^M  
+ E^A_{\al}\del_{\kappa}\th^{\al}  \nn \\
&\equiv& H^A_{~B}\del_{\kappa}X^M e_M^B + E^A_{\al}\del_{\kappa}\th^{\al}\,. 
\label{eqn:deltaX}
\end{eqnarray}
Here, in order to make the structure clear, let us
introduce the following abbreviations:
\begin{eqnarray}
\del_{\kappa}X^M e_M^A \equiv \del x^A\,, \qquad 
\del_{\kappa}\th E^A \equiv \del \th^A\,, 
\end{eqnarray}
and then (\ref{eqn:deltaX}) is written as 
\begin{eqnarray}
\del x^A = H^A_{~B}\del x^B + \del\th^A\,. 
\end{eqnarray}
By using this equation recursively, we can derive the following
expression: 
\begin{eqnarray}
\delta x^A = \left(H^{15} + \dots + 1\right)^{A}_{~B}\del\theta^B\,. 
\label{se}
\end{eqnarray}
Now let us evaluate each of terms in the r.\ h.\ s.\ of (\ref{se}) and 
show that all of them are zero in the case that $A$ is a Dirichlet
direction.
Noting that
\begin{eqnarray}
\delta\theta^B=
E_{\al}^B \del_{\kappa}\th^{\al} = i\bar{\th}\Gamma^B 
\left(\frac{\sinh(\mathcal{M}/2)}{\mathcal{M}/2}\right)^2 \del_{\kappa}\th 
\left\{
\begin{array}{l}
\neq 0 \quad 
\mbox{$B$: Neumann} 
\\
= 0 \quad 
\mbox{$B$: Dirichlet} 
\end{array}
\right.\,,
\label{im}
\end{eqnarray}
the equation (\ref{ho}) becomes 
\begin{eqnarray}
\delta x^{\underline{A}}=\left(H^{15} + \dots + 1\right)^{\underline{A}}{}_{\overline{B}}
\delta\theta^{\overline{B}}=0\,.
\label{eqn:delH}
\end{eqnarray}
The zero-th order term with respect to $H$ is obviously zero 
because $\delta^{\underline{A}}_{~\overline{B}}=0$\,. 
The first order term $H^{\underline{A}}_{~\overline{B}}$ is written as 
\begin{eqnarray}
 H^{\underline{A}}_{~\overline{B}}&=&
  -i\bar{\th}\Gamma^{\underline{A}}
\left(\frac{\sinh(\mathcal{M}/2)}{\mathcal{M}/2}\right)^2 
[D\th]_M e^M_{\overline{B}} \nn \\
&=&  - i\bar{\th}\Gamma^{\underline{A}}P^{\mp}
\left(\frac{\sinh(\mathcal{M}/2)}{\mathcal{M}/2}\right)^2 P^{\mp} 
[D\th]_M e^M_{\overline{B}}\,. 
\end{eqnarray}
We can rewrite furthermore $P^{\mp}[D\th]_M e^M_{\overline{B}}$ as 
\begin{eqnarray}
P^{\mp}[D\th]_M e^M_{\overline{B}} &=& P^{\mp}
\left(
\frac{\lambda}{2}e^C_M\widehat{\Gamma}_Ci\sigma_2\th + \frac{1}{4}\omega_M^{CD} \Gamma_{CD}\th 
\right) e^M_{\overline{B}}  \nn \\
&=& \frac{\lambda}{2}\widehat{\Gamma}_{\overline{B}}i\sigma_2P^\mp \th
 + \frac{1}{4}\omega_{\overline{B}}^{\overline{C}\underline{D}}\Gamma_{\overline{C}
\underline{D}}P^\pm\th\,. 
\end{eqnarray}
The first term in the last line vanishes for $P^\pm\theta=\theta$\,.
Because
the components of spin connection
vanish at the origin, i.e. $\omega_{\overline{B}}^{\overline{C}
\underline{D}} = 0$\,,  
the term $H^{\underline{A}}{}_{\overline{B}}$ should vanish at the origin. 
Using this fact, we find that
\begin{eqnarray}
H^{\underline{A}}{}_{{C_1}} H^{{C_1}}{}_{{C_2}}\cdots 
H^{C_n}{}_{\overline{B}}=
H^{\underline{A}}{}_{\underline{C_1}} H^{\underline{C_1}}{}_{\underline{C_2}}
 \cdots H^{\underline{C_n}}{}_{\overline{B}}=0\,.
\end{eqnarray}
Thus, we have shown the useful identity (\ref{eqn:delH}), and so (\ref{ho})\,. 

Now let us return to the proof of (\ref{key2}) and consider 
\begin{eqnarray}
P^{\mp}\del_{\kappa}Z^{\widehat{M}}
E_{\widehat{M}}^{\al} = 
P^{\mp}\left(\del_{\kappa}X^M E_M^{\al} + \del_{\kappa}\th^{\beta}
E_{\beta}^{\al}\right)\,,
\label{hon}
\end{eqnarray}
for the boundary condition $\th = P^{\pm}\th$\,. 
First, we shall consider the first term in r.\ h.\ s.\ of (\ref{hon})\,, which 
can be rewritten as 
\begin{eqnarray}
&& P^{\mp}\frac{\sinh\mathcal{M}}{\mathcal{M}}[D\th]_M\del_{\kappa}X^M 
= P^{\mp}\frac{\sinh\mathcal{M}}{\mathcal{M}}P^{\mp}[D\th]_{\overline{A}}
e^{\overline{A}}_{M}\del_{\kappa}X^M \nn \\
&& \qquad \qquad  
= P^{\mp}\frac{\sinh\mathcal{M}}{\mathcal{M}}P^{\mp}
\left(
\frac{\lambda}{2}\widehat{\Gamma}_{\overline{A}}i\sigma_2\th 
+ \frac{1}{4}\omega^{BC}_{\overline{A}}\Gamma_{BC}\th 
\right) e^{\overline{A}}_M\del_{\kappa}X^M\,. \label{hon2}
\end{eqnarray}
In the first line of (\ref{hon2})\,, we have used the relations,
(\ref{PMP}) and (\ref{ho})\,.  The first term in the last line of
(\ref{hon2}) always vanishes for the condition $\th = P^{\pm}\th$\,. The
second term is also zero at the origin by using $\th =
P^{\pm}\th$\,. Hence we have seen that the first term in
(\ref{hon}) vanishes.  Furthermore, we can see that the second term in
(\ref{hon}) also vanishes from the relation:
\begin{eqnarray}
P^{\mp}\frac{\sinh\mathcal{M}}{\mathcal{M}}\del_{\kappa}\th = 
P^{\mp}\frac{\sinh\mathcal{M}}{\mathcal{M}}P^{\mp}\del_{\kappa}\th = 0\,,
\end{eqnarray}
because of $\th = P^{\pm}\th$\,. 
Thus, we have shown that the second key relation (\ref{key2})\,. 

Finally, we would like to comment on the physical implication of the
condition (\ref{key2})\,. When we consider the flat limit ($\lambda
\rightarrow 0$)\,, (\ref{key2}) is reduced to the 1/2 SUSY 
condition:
\begin{eqnarray}
P^{\mp}\del_{\kappa}\th = 0\,. 
\end{eqnarray}
Hence (\ref{key2}) may be a generalization of 
projection condition to the AdS$_5\times S^5$ case.

\section{Conclusion and Discussion} 

We have shown that higher order surface terms with respect to $\th$\,,
which come from the $\kappa$-variation, do not affect the classification
obtained in the fourth order analysis.  This proof completes our
previous classification at the fourth order.  As a matter of course, our
proof is obviously applicable to the full order analysis of D-branes in
the type IIB string on the pp-wave background.  But the validity of
D-brane classification in the pp-wave should be obvious via the Penrose
limit \cite{P,BFHP2,HKS:10} of D-branes on the $AdS_5\times S^5$\,.
Hence, we may say that the fourth order analysis in the covariant
formulation is sufficient to classify the possible configurations of
D-branes.

In this paper, we have considered D-branes of open F- and D-strings,
and presented a simple prescription
for the vanishing conditions of the $\kappa$-variation
surface terms.
Our scenario can be extended to D-branes of an open D$p$-brane
in an obvious way.
We reserve this issue for the next publication.
Also, it is not quite trivial
in the case of open supermembrane on the pp-wave and AdS$_{4/7}\times
S^{7/4}$ backgrounds because of the dimensionality. 
We will report on these cases in another place soon 
\cite{future}.

\section*{Acknowledgments}

We would like to thank M.~Hatsuda.  
This paper is supported by the 21 COE program
"Constitution of wide-angle mathematical basis focused on knots".

\end{document}